# Comment on "Uncovering the Complex Behavior of Hydrogen in Cu₂O"

In a recent Letter [1] Scanlon and Watson (SW) reported their first principles results on hydrogen in $Cu_2O$. Their main conclusions are: (1) an interstitial H in $Cu_2O$ prefers to occupy the tetrahedral site ($H_i^{tet}$), which is coordinated with four Cu cations, in all three charge states (+1, neutral, and -1); (2) H will bind with a Cu vacancy and form an electrically active H-$V_{Cu}$ defect complex, which is amphoteric with (+/0) and (0/−) transition levels at $E_v + 0.1$ and $E_v + 1.1$ eV, respectively. However, these two conclusions contradict two generally observed behaviors of H in oxides: (i) cationic H usually binds with an O atom, forming a single O-H bond, while the anionic H usually binds with cations with multi-coordination; (ii) H usually passivates cation vacancies in oxides. In this Comment, we explicitly show that with charge state +1, H prefers to bind with a single O anion rather than with four Cu cations and that H-$V_{Cu}$ does not induce any defect levels inside the band gap. Our results were obtained by using the similar computational method as used in Ref. 1, i.e., hybrid density functional method as implemented in VASP.

*Hydrogen interstitial*: We show in Fig. 1 that with charge state +1, $H_i$ with a single O-H bond (denoted $H_i^{AB2}$ in Ref. 1) is more stable than $H_i^{tet}$ (with four Cu-H bonds) by 0.29 eV, opposite to SW. On the other hand, with charge state -1, $H_i^{tet}$ is the most stable structure with an anionic H binding with four Cu cations. Our results are consistent with general H chemistry in oxides. However, it is interesting to note that the metastable $H_i^{tet,+}$ is a donor even though H binds four Cu cations; thus $H_i^{tet}$ is amphoteric as shown in Fig. 1. SW described the H in $H_i^{tet}$ as "quasi-atomic" with multiple charge states. This is, however, incorrect. Our analysis shows that *H is anionic in all three charge states of* $H_i^{tet}$, as expected from the general H chemistry in oxides. The flexibility of $H_i^{tet}$ taking various charge states is due to a change in the charge state of the four Cu atoms that bind H, since the oxidation state of Cu usually ranges from +1 to +2. The hybridization of the H 1s and the Cu 3d states in $H_i^{tet}$ produces a bonding state of mainly H character below the valence band and an antibonding state of mainly Cu 3d character inside the band gap. The occupation number of the Cu gap state determines the charge state of $H_i^{tet}$.

*H-$V_{Cu}$ complex*: In this complex, $H_i$ forms a strong O-H bond with one of the two O atoms near $V_{Cu}$. A neutral H-$V_{Cu}$ has no defect states in the band gap and therefore should not be amphoteric with the (0/+) and (0/-) transition levels reported in Ref. 1. Our attempts to calculate H-$V_{Cu}$ in both +1 and -1 charge states lead to a hole at the VBM and an electron at the CBM, respectively, confirming that H-$V_{Cu}$ is electrically inactive. However, SW observed a deep level (located near midgap at $E_v + 1.0$ eV), which they assigned to H-$V_{Cu}$. This state was claimed to be responsible for the (0/-) transition level of H-$V_{Cu}$. Surprisingly, SW also stated that the observed deep state is a highly delocalized Cu d state that spreads throughout the supercell. This contradicts the fact that a deep level has to be highly localized with localization related to the distance from the band edge. We did not observe this state in our calculations.

Besides $H_i$ and H-$V_{Cu}$, SW also claimed that substitutional H ($H_O$) is amphoteric because it induces a deep gap state and can have +1, 0, or -1 charge state. The result that $H_O^+$ can trap one or two additional electrons is abnormal, because, in $H_O^-$, the H⁻ anion is fully reduced and reduction of its neighboring Cu⁺ cations is not expected in an oxide. We find that the defect state SW referred to is not a localized deep level but a very dispersive defect band of Cu-H antibonding character. Our calculations show that this band is above the CBM at $\Gamma$ point but falls below the CBM at some k points, which might be sampled by SW in the Brillouin zone integration. The large dispersion of the defect band is a result of the small 48-atom supercell used in the calculation. Therefore, the (+/0) and (0/-) levels of $H_O$ calculated by SW are incorrect (far from convergence with respect to the supercell size). In a large supercell calculation, which is unfortunately too expensive, the Cu-H antibonding state is expected to become a resonant state above the CBM, resulting in $H_O$ being a shallow donor as seen in many oxides.

Work at ORNL was supported by DOE-BES Materials Sciences and Engineering Division; work in Thailand was supported by TRF (RTA5280009).

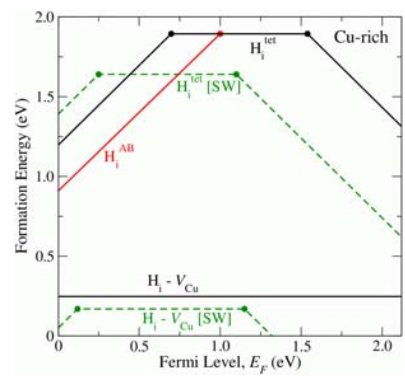

FIG. 1 (color online). Formation energies of $H_i$ and $H_i$-$V_{Cu}$ in $Cu_2O$. Dashed lines show the results in Ref. 1.

K. Biswas,[1] M. -H. Du,[1] J. T-Thienprasert,[2] S. Limpijumnong,[1,4] and D. J. Singh[1]

[1] Oak Ridge National Laboratory, Oak Ridge, TN 37831, USA

[2] Department of Physics, Kasetsart University, Bangkok 10900, Thailand

[3] Thailand Center of Excellence in Physics (ThEP Center), Commission on Higher Education, Bangkok 10400, Thailand

[4] School of Physics, Suranaree University of Technology and Synchrotron Light Research Institute, Nakhon Ratchasima 30000, Thailand

[1] D.O. Scanlon and G.W. Watson, Phys. Rev. Lett. **106**, 186403 (2011).